\documentclass[epj,pacs]{svjour}

\usepackage{latexsym}
\usepackage{amsmath}
\usepackage{amssymb}
\usepackage{indentfirst}
\usepackage{color}
\usepackage{graphicx}
\usepackage{bm}
\usepackage{maybemath}
\usepackage{revsymb4-1}

\RequirePackage{fix-cm}
\RequirePackage{graphicx}
\RequirePackage{latexsym}
\RequirePackage[numbers,sort&compress]{natbib}
\RequirePackage[colorlinks,citecolor=blue,urlcolor=blue,linkcolor=blue]{hyperref}

\newcommand{\dd}{\mathrm{d}}
\newcommand{\ee}{\mathrm{e}}

\journalname{EPJC}

\begin{document}

\def\asq#1{\textbf{..(??. #1 .??)..} }          % Andrey's question or comment

%
% Bound--free pair production in ultra--relativistic ion collisions at the LHC collider
%
\title{Bound--free pair production in ultra--relativistic ion collisions at the LHC collider: Analytic approach to the total and differential cross sections}

\titlerunning{Bound--free pair production at LHC}

\author{A.~N.~Artemyev\inst{1,2} \and
U.~D.~Jentschura\inst{3} \and
V.~G.~Serbo\inst{4,5,}\thanks{e--mail: serbo@math.nsc.ru} \and
A.~Surzhykov\inst{1,2}}

\authorrunning{Artemyev, Jentschura, Serbo, Surzhykov}

%
% Affiliations
%

\institute{Physikalisches Institut der Universit\"at Heidelberg,
Philosophenweg 12, 69120 Heidelberg, Germany
\and
GSI Helmholtzzentrum f\"ur Schwerionenforschung GmbH,
Planckstra\ss{}e 1, 64291 Darmstadt, Germany
\and
Department of Physics, Missouri University of Science and Technology,
Rolla, MO 65409--0640, USA
\and
Sobolev Institute of Mathematics, Koptyug avenue 4,
630090, Novosibirsk, Russia
\and
Novosibirsk State University, Pirogova Street 2,
630090, Novosibirsk, Russia}

\date{Received: date / Accepted: date}

%
% Abstract
%
\abstract{A theoretical investigation of the bound-free electron-positron pair production
in relativistic heavy ion collisions is presented.  Special attention is paid
to the positrons emitted under large angles with respect to the beam direction.
The measurement of these positrons in coincidence with the down--charged ions is in principle
feasible by LHC experiments. In order to provide reliable estimates
for such measurements, we employ the equivalent photon approximation
together with the Sauter approach and derive simple analytic expressions for the
differential pair--production cross section, which compare favorably
to the results of available numerical calculations. Based on the analytic expressions,
detailed calculations are performed for collisions of bare Pb$^{82+}$ ions,
taking typical experimental conditions of the LHC experiments into account.
We find that the expected
count rate strongly depends on the experimental parameters
and may be significantly enhanced by increasing the positron--detector acceptance cone.\\
{\bf Key words.}~Bound--free pair production,
Relativistic heavy--ion collisions,
Virtual photons\\
{\bf PACS}~25.75.Dw, 12.20.Ds, 25.75.-q, 03.65.Pm}

\maketitle

%
% Introduction
%
\section{Introduction}
\label{Sec_Introduction}

Owing to recent advances in heavy--ion accelerators, an
increasing interest arises in exploring electromagnetic
processes accompanying ion collisions.
One of the dominant processes is electron--positron pair production;
pertinent cross sections are large.
For ultra--relativistic collisions of two bare lead ions
(Pb$^{82+}$), for example, the cross section of the creation of a
\textit{free} $e^+ e^-$ pair may reach hundreds of kilobarns
\cite{BuG75,IvS99,LeM02,LeM09,Bau07}. Both
free--free as well as~\textit{bound--free} pair production can take
place in which the electron is captured by one of the projectiles
resulting, thus, in the formation of a hydrogen--like ion,
\begin{equation}
\label{eq_process}
Z_1 + Z_2 \to Z_1 + e^+ + (Z_2+e^-) \,,
\end{equation}
where the bound system is denoted by round brackets.
Even though this (bound--free) process is usually orders of
magnitude less probable than free--free pair production, its
investigation is of great importance not only for a better
understanding of the physics of extraordinary strong
electromagnetic fields but also for the development and operation
of novel collider facilities. In free--free
$e^+e^-$ production, the scattered nuclei lose only a very small fraction
of their energy and acquire tiny scattering angles, and thus do not
leave the beam. In contrast, if one of the colliding ions captures
an electron [cf.~Eq.~(\ref{eq_process})], it changes its charge
state and is bent out from the beam. The
corresponding cross section is rather large, about $100$ barn, and
the reaction (\ref{eq_process}) is one of the important processes which
limit the luminosity of colliders. Besides, the secondary
beams of down-charged ions emerging from the collision point hit
beam-pipe and deposit a considerable portion of energy at a
small spot, which may in turn lead to the quenching of superconducting
magnets~\cite{Bau02,JoB05,Bru07,BrB09}.

Because of its fundamental and practical importance, the
bound--free pair production has been in the focus of intense
research over the past years. A series of experiments have been
performed, for example, at the CERN Super Proton Synchrotron (SPS)
to analyze the total cross sections of the process for
ultra--relativistic collisions of highly--charged Pb~ions with
solid--state and gas targets \cite{Kra98,Kra01}. These
experimental findings are currently understood based on
theoretical predictions of
relativistic Dirac theory \cite{BaR96,VoN08}. In contrast to the
\textit{total} rates, much less attention has been paid until now
to the \textit{differential} bound-free pair production cross sections.
However, the angle--resolved analysis of the positron emission is
of definite interest since it allows to probe a parameter range
(of the $e^+ e^-$ process) which is otherwise not accessible in
total cross section studies. Namely, while the total probabilities
arise from a region where the transverse momentum of the
produced positron is small, $p_{+\perp} \lesssim m_e$, large
values of $p_{+\perp}$ may significantly contribute to the
angle--differential rates.

An understanding of the differential pair production properties is
also significant for the analysis of future experiments at the LHC facility.
In these experiments,
the down--charged ions can be measured \textit{in coincidence}
with the emitted positrons in order to reduce background events and
clearly distinguish the process~(\ref{eq_process}). However, since
the central detector is likely to be placed at a rather large
angle with respect to the beam direction it will only detect
those positrons, whose transverse momenta are much greater
than the electron mass, $p_{+\perp} \gg m_e$. An
estimate of the yield of these positrons is highly desirable for
estimating the feasibility of future measurements;
the task has not been fully tackled until now
to the best of our knowledge.

In this contribution, therefore, we present a theoretical study of bound--free
pair production with a special emphasis on the differential cross sections. The
exact computations of these cross sections for the ultra--relativistic ion
collisions and large positron momenta are very laborious, and we shall develop
an \textit{approximate} method that allows for an analytic treatment.  Before
discussing the derivation of the analytic expressions, we introduce in
Section~\ref{Sec_parameters} the notation used in the manuscript and estimate
typical values of kinematic parameters corresponding to the LHC experiments.
By using these estimates and the equivalent photon approximation (EPA) we
express in Sect.~\ref{Sub_sec_EPA} the differential probabilities of the
process~(\ref{eq_process}) in terms of the pair \textit{photo}--production
cross sections. The evaluation of these cross sections within the framework of
the Sauter approximation is discussed in
Sects.~\ref{Sub_sec_SLA_differential} and~\ref{Sub_sec_cross_sections}.  The
validity of our EPA--Sauter model is tested numerically in
Sects.~\ref{Sub_sec_total_calculations}
and~\ref{Sub_sec_differential_calculations} where we calculate the total as
well as differential pair production rates and compare them with the
predictions of an exact relativistic theory. Based on the results of this test,
we employ the EPA--Sauter approach in order to estimate the positron yield for
ultra--relativistic collisions of bare Pb$^{82+}$ as relevant to the LHC
studies. In Sect.~\ref{Sub_sec_differential_calculations} we focus on two
experimental scenarios: In the first scenario, only positrons with very large
transverse momenta, $p_{+\perp}\sim 1$~GeV are ``seen'' by the detectors, while
much smaller momenta, $p_{+\perp} \sim 0.05$~GeV, are considered in the second
case.  We find that the reduction of the (accepted) minimum momentum
$p_{+\perp}$ leads to an increase of the positron count rate from just a single
event in 67 days to about $10$~events per hour. We conclude
with a brief summary in Sect.~\ref{Sec_summary}.
Through the paper, we use relativistic units with $c= \hbar = 1$
and $\alpha \approx 1/137$.

%
% Notations and kinematic parameters
%
\section{Notations and kinematic parameters}
\label{Sec_parameters}

Let us first recall basic notations and assumptions used throughout this
paper. The initial state of the overall system is given by two bare ions of
charges $Z_1$ = $Z_2$ = $Z$ and masses $M_1$=$M_2$=$M$, moving towards each
other with 4--momenta $P_{1,2} = (E_{1,2}, {\vec  P}_{1,2})$ and
corresponding Lorentz--factors $\gamma_1$ = $\gamma_2$ = $\gamma$. Being
defined in the \textit{collider} frame, these kinematic parameters are
convenient for the description of experimental results. However, the
theoretical study of atomic processes accompanying ion--ion collisions is most
conveniently done in the rest frame of that particular ion which finally becomes
hydrogen--like. For definiteness, we shall assume electron to be captured
by the ``second'' nucleus in whose rest frame the ``first'' nucleus moves with
a Lorentz factor $\gamma_L = 2\gamma^2 - 1$. By adopting $\gamma$ = 1500, which
is the typical value for LHC experiments on Pb--Pb ($Z_1$ = $Z_2$ = 82)
collisions, we find $\gamma_L = 4.5 \, \cdot 10^{6}$. This value is used
in all numerical estimates below.

Bound--free pair production in energetic ion
collisions can be uniquely detected experimentally by measuring
the down--charged ion \textit{in coincidence} with the emitted
positron. In the set--up of the typical LHC experimental arrangement, the
positron detector is placed at rather large angles with respect to
the incident beam direction. Therefore, only positrons with a
transverse momentum
\begin{equation}
\label{eq_transverse_momentum_min}
p_{+\perp} \geq  p_{\min} \gg m
\end{equation}
can be observed in a particular experiment. In
Eq.~(\ref{eq_transverse_momentum_min}), $m \equiv m_e$ is the
electron mass and momenta are given in the collider frame. In
the same frame, we define the positron's rapidity:
\begin{equation}
\label{eq_rapidity_definition} y_+ = \frac{1}{2}
\ln\frac{\varepsilon_++p_{+z}}{\varepsilon_+-p_{+z}} \approx
-\ln\left[ \tan\left(\tfrac12 \, \theta_+\right)\right] \, ,
\end{equation}
where $\varepsilon_+$ and $p_{+z}$ denote the energy and
longitudinal (along the beam axis) momentum of the positron and
$\theta_+$ its emission angle. These quantities are directly
related to the transverse momentum, as follows,
\begin{equation}
\label{eq_energy_longitudinal_momentum}
\varepsilon_+ \approx \frac{p_{+\perp}}{\sin{\theta_+}} \,, \;\;
p_{+z} = \frac{p_{+\perp}}{\tan{\theta_+}} \,.
\end{equation}
Equation~(\ref{eq_transverse_momentum_min}) immediately implies
restrictions both on the rapidity,
\begin{equation}
   \label{eq_rapidity_restriction}
   -y_{\min} \leq y_+ \leq y_{\min} \, ,
\end{equation}
as well as on the positron angle,
\begin{equation}
\label{eq_angle_restriction}
\theta_{\min} \leq \theta_+ \leq \pi - \theta_{\min}\,,
\qquad
\theta_{\min} = 2 \, \arctan\left( \ee^{-y_{\min}} \right)\,.
\end{equation}

In bound-free pair production,
certain limitations are imposed on the electron's kinematic
parameters. As the electron is produced in bound
ionic states, its relativistic Lorentz factor and scattering angle
match those of the ``second'' nucleus. Indeed, for
ultra--relativistic collisions of heavy high--$Z$ ions, the
corresponding scattering angle should be small,
\begin{equation}
\label{eq_nuclaer_scattering_angle}
\theta_2 \lesssim \frac{p_{+\perp}}{E_2} =
\frac{p_{+\perp}}{\gamma \, M} \sim 10^{-5} \,.
\end{equation}
Here, all parameters are given in the collider frame. It
follows from Eq.~(\ref{eq_nuclaer_scattering_angle}) and the
previous discussion, that the electron's momentum is almost
parallel to the initial direction of propagation of the ``second''
nucleus, chosen as the $z$--axis:
\begin{equation}
\label{eq_electron_momentum_energy}
p_{-\perp} \approx 0 \,,
\qquad
p_{-z} \approx \varepsilon_- = m \gamma \, .
\end{equation}
Using these expressions and
Eqs.~(\ref{eq_transverse_momentum_min})--(\ref{eq_energy_longitudinal_momentum})
for positron emission, we can find the four--momentum
$p = (\varepsilon, {\bf p}) =
(\varepsilon_+ + \varepsilon_-, {\bf p}_+ + {\bf p}_-)$ and the invariant mass
\begin{eqnarray}
\label{eq_invariant_mass}
W &=& \sqrt{p^2}\approx \sqrt{2p_+ p_-} =
\sqrt{2p_{+\perp}\, m\gamma \,
\tan\left(\tfrac12 \,\theta_+\right)} \nonumber \\[0.2cm]
&\geq& \sqrt{2 \, p_{\min}\, m\gamma \, \tan\left(\tfrac12 \, \theta_{\min}\right)} \,
\end{eqnarray}
of the final $e^+ e^-$ system.

% Figure 1
\begin{figure}[t]
\begin{center}
     \includegraphics[width=7cm]{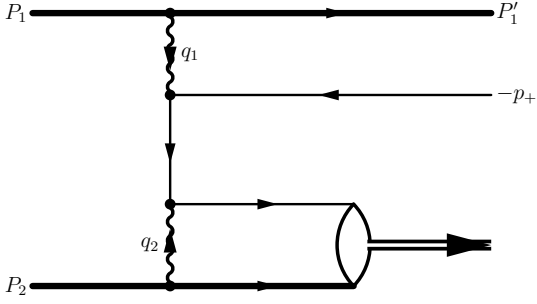}
\end{center}
\caption{Schematic Feynman diagram for the bound--free pair
production in a heavy--nucleus collision. Heavy colliding nuclei
are represented by thick lines while thinner lines correspond to
light fermions (electron and positron).
The trajectory of the produced electron is aligned with that of the
second nucleus.}
 \label{Fig1}
\end{figure}

Our theoretical analysis of the bound--free pair production is
based on a quantum electrodynamic (QED) approach.
In lowest order, the process (\ref{eq_process}) is described
by the diagram in Fig.~\ref{Fig1}. The
exchange of two virtual photons, with four momenta $q_{1,2} =
(\omega_{1,2}, {\bf q}_{1,2})$, is required to produce the
$e^+ e^-$ pair. For the present study, it important to know the
\textit{virtuality} $Q^2$ of these photons which measures
how far they are off the mass shells. For the first gamma quantum
with a 4--momentum $q_1 = P_1 - P'_1$, for example, we write
\begin{equation}
\label{eq_photon_virtuality}
Q^2  \equiv  -q^2_1 \; = \; -\omega_1^2+q^2_{1\perp}+q^2_{1z} \;
\approx \; q_{1\perp}^2+\frac{\omega_1^2}{\gamma_1^2} \, ,
\end{equation}
where we use the fact that
\begin{subequations}
\label{eq_photon_components}
\begin{align}
\omega_1 =& \; E_1 - E'_1\,,\;
\\[2ex]
q_{1z} \approx & \;
-E_1+\frac{M^2}{2E_1}+E'_1-\frac{M^2+q^2_{1\perp}}{2E'_1}\,.
\end{align}
\end{subequations}
The virtual photons are ``emitted'' by counter--propagating
nuclei, and we finally obtain
\begin{equation}
\label{eq_photon_momenta}
q_{1z} \approx -\omega_1 \,, \qquad q_{2z}\approx +\omega_2 \,.
\end{equation}
Together with the four--momentum conservation relations
$\omega_1 + \omega_2=\varepsilon$ and $q_{1z} + q_{2z} = p_z$ and by
employing Eqs.~(\ref{eq_energy_longitudinal_momentum}) and (\ref{eq_electron_momentum_energy}),
this allows us to derive the energy of the first virtual photon,
\begin{eqnarray}
\label{eq_energy_first_photon}
\omega_1 &=& \frac{1}{2} \left(\varepsilon -p_z\right) = \frac{1}{2} \left(\varepsilon_{+} - p_{+z} + \varepsilon_{-} - p_{-z} \right)
\nonumber \\[0.2cm]
&=& \frac{1}{2} \left(\frac{p_{+\perp}}{\sin{\theta_+}}- \frac{p_{+\perp}}{\tan{\theta_+}}\right) = \frac{1}{2} p_{+\perp} \,
\tan\left(\tfrac12 \, \theta_+\right) \,,
\end{eqnarray}
as a function of the transverse momentum and emission angle of
positron.
Employing the restrictions given in Eqs.~(\ref{eq_transverse_momentum_min})
and~(\ref{eq_angle_restriction}) on these quantities, the
\textit{minimum} allowed value of $\omega_1$ can be found,
\begin{equation}
\label{eq_energy_first_photon_minimum}
\min\{\omega_1\} = \frac 12 \, p_{\min} \,
\tan\left(\tfrac12 \, \theta_{\min} \right) \,.
\end{equation}

Before we proceed to the theoretical treatment of the $e^+ e^-$
pair production, let us first compute some typical values
for the basic kinematic parameters from above. Assumptions about the
\textit{magnitude} of these values will be required later
in order to
perform a number of approximations that greatly simplify the
calculations of the cross sections. As we have seen already, the
ranges of positron as well as virtual photon parameters are
crucially depend on the particular experimental set--up.
Here, two distinct experimental ``scenarios'' are considered.

In the {\it first scenario}, which relies on a typical LHC--detector
acceptance, the transverse momentum and
rapidity of the positrons are restricted by the conditions:
\begin{equation}
\label{eq_scenario_A0}
p_{\min} = 1\;\mbox{GeV}\,,\qquad y_{\min}=1\,.
\end{equation}
For these parameters, we find
\begin{subequations}
\label{eq_scenario_A}
\begin{align}
\theta_{\min} =& \; 40^\circ \,, \\[0.77ex]
W \geq & \; 0.75 \;\mbox{GeV} \,,  \\[0.77ex]
\min\{\omega_1\}=& \; 0.18 \;\mbox{GeV} \,,
\end{align}
\end{subequations}
for the minimum values of the positron emission angle
$\theta_{\min}$, the virtual photon energy $\min\{\omega_1\}$ as well as
for the $e^+ e^-$ invariant mass $W$, respectively.

In the {\it second scenario}, we assume smaller
transverse momenta of the positron and a greater rapidity:
\begin{equation}
\label{eq_scenario_B0}
p_{\min} = 0.05\;\mbox{GeV}\,,\qquad y_{\min}=1.5\,,
\end{equation}
which results in significantly different parameters:
\begin{subequations}
\label{eq_scenario_B}
\begin{align}
\theta_{\min} =& \; 25^\circ \,, \\[0.77ex]
W \geq & \; 0.13 \;\mbox{GeV}, \,, \\[0.77ex]
\min\{\omega_1\} =& \; 5.6 \;\mbox{MeV} \,.
\end{align}
\end{subequations}
As we see below in
Section~\ref{Sub_sec_differential_calculations}, the
prospects for a unique detection of bound-free pair production
seem much more favorable for the
parameters listed in Eq.~(\ref{eq_scenario_B}) than for the
parameters in scenario~(\ref{eq_scenario_A}).

%
% Theoretical background
%
\section{Theoretical background}
\label{Sec_Theory}

\subsection{Equivalent photon approximation}
\label{Sub_sec_EPA}

Since its introduction by Fermi \cite{Fer24} and further
development by von Weizs\"acker and Williams, the equivalent
photon approximation (EPA) has been successfully applied to the
description of a large number of electromagnetic processes induced
in collisions of charged particles. Usually, one physically
justifies this method based on the observation that the
electromagnetic field of a fast moving charge becomes almost
transverse, and the electric and magnetic fields both have about
equal strengths~\cite{RaM90}. In the observer (laboratory)
frame, therefore, the projectile's fields can be \textit{seen} as
those of the pulse of a plane, linearly polarized wave. The
frequency spectrum of such a pulse is calculated within a
semiclassical model and used in order to evaluate
collision--induced cross sections \cite{EiM95}.

Even though the conventional EPA is found very useful for the
theoretical analysis of ion collisions, here we will employ an
alternative approach which---in the ultra--relativistic
domain---appears to be more transparent and rigorous and allows for a
simple estimate of its accuracy. Our approach exploits the EPA
as an \textit{approximate} method for calculating a Feynman
diagram for the corresponding process and uses the fact that
the virtual photons in the diagram are close to the mass
shell---for details, see the review~\cite{BuG75}. The diagram
for bound--free pair production (\ref{eq_process}) is
displayed in Fig.~\ref{Fig1}, where the two thick lines represent
the colliding nuclei and the double--line arrow just refers to the
residual hydrogen--like ion. Inspecting this diagram, we may
interpret the bound--free pair production as being due to the
interaction of the second (lower) nucleus with the virtual (or
equivalent) photon with the energy $\omega_1$ and virtuality $Q^2
\equiv - q_1^2$ emitted by the first nucleus. Thus, the
theoretical analysis of the $e^+e^-$ creation in energetic
heavy--ion collisions can be traced back to the virtual process:
\begin{equation}
\label{eq_EPA_pair_production}
\gamma^* + Z_2 \to e^+ + (Z_2 + e^-)_{1s} \, ,
\end{equation}
where the electron is captured into the ground ionic state. The
differential cross section d$\sigma_{ZZ}$ of the process
(\ref{eq_process}) can be expressed, therefore, in terms of the
cross sections $\sigma^T_{\gamma* Z}$ and $\sigma^S_{\gamma* Z}$
for the (virtual) pair production (\ref{eq_EPA_pair_production})
as:
\begin{equation}
\label{eq_EPA_cross_section}
{\dd}\sigma_{ZZ} = {\dd}n_T\, {\dd}\sigma^T_{\gamma* Z} +
{\dd}n_S\, {\dd}\sigma^S_{\gamma* Z} \, .
\end{equation}
In this expression, we denote by ${\dd}n_T$ and ${\dd}n_S$
the numbers of the equivalent transverse $T$ and scalar (or
longitudinal) $S$ photons and, moreover, neglect the interaction
between the emitted positron and the first nucleus.

As seen from Eq.~(\ref{eq_EPA_cross_section}), any further
evaluation of ${\dd}\sigma_{ZZ}$ requires the knowledge of
both of the pair photo--production cross sections
${\dd}\sigma^{T,S}_{\gamma* Z}$
and infinitesimal numbers ${\dd}n_{T,S}$ of equivalent
photons. Below we will evaluate these quantities in the
rest frame of the second ion, which finally
constitutes a hydrogen--like ion. The energy of the equivalent
photon, emitted by the first nucleus (cf. Fig.~\ref{Fig1}), is
given in such a frame as:
\begin{equation}
\label{eq_photon_energy_proj_frame}
\omega_L = \frac{q_1P_2}{M} \approx 2\gamma \omega_1\,.
\end{equation}
For the case of an LHC experiment, the minimum value of this energy,
$\min\{\omega_L\}=2\gamma \min\{\omega_1\}$, is huge: it
amounts to about $550$~GeV
and $17$~GeV for the first (\ref{eq_scenario_A}) and second (\ref{eq_scenario_B})
scenarios, respectively.

In ultra--relativistic nuclear collisions, the main contribution to the pair
production cross section (\ref{eq_EPA_cross_section}) is given by a wide range
of the photon virtualities (\ref{eq_photon_virtuality}) ranging from the
small minimum value
\begin{equation}
\label{eq_virtuality_minimum}
Q^2_{\min} = \frac{\omega_L^2}{\gamma_L^2} \, ,
\end{equation}
to some $Q^2_{\max}$ which is derived below from the
analysis of $\dd n_{T,S}$ and the differential cross sections
$\dd \sigma^{T,S}_{\gamma* Z}$. In such a region, the numbers of
equivalent photons are given by (see Appendix D of Ref.~\cite{BuG75}):
\begin{eqnarray}
\label{eq_photon_number_estimate}
\dd n_S(\omega_L,Q^2) &\sim& \dd n_T(\omega_L,Q^2) \nonumber \\[0.2cm]
& & \hspace*{-2cm} =
\frac{Z_1^2\alpha}{\pi} \,
\frac{\dd \omega_L}{\omega_L} \,
\frac{\dd Q^2}{Q^2}\,
\left( 1 - \frac{Q^2_{\min}}{Q^2} \right)\,
F^2(Q^2) \,,
\end{eqnarray}
where $F(Q^2)$ is the form factor of the first nucleus.

Integrating Eq.~(\ref{eq_photon_number_estimate}) over the
interval $[Q^2_{\min}, Q^2_{\max}]$, we find the overall numbers
$\dd n_S(\omega_L)$ and  $\dd n_T(\omega_L)$ of the scalar and
transverse virtual photons which contribute to the production
process (\ref{eq_process}). For the $dn_T(\omega_L)$ we obtain:
\begin{eqnarray}
\label{eq_photon_number_integral}
\dd n_T(\omega_L) &=&
\frac{Z_1^2\alpha}{\pi} \, \frac{\dd \omega_L}{\omega_L}
\nonumber\\
&\times&
\int_{Q^2_{\min}}^{Q^2_{\max}} \frac{\dd Q^2}{Q^2}\,
\left( 1 - \frac{Q^2_{\min}}{Q^2} \right)\,F^2(Q^2)\,.
\end{eqnarray}
The evaluation of this integral is discussed in detail in Ref.~\cite{JeS09} and
employs the $Q^2$-behavior of the (square of) the nuclear form factor. For
example, the function $F^2(Q^2)$ drops quickly when the virtuality of the
photon becomes greater than the squared inverse electromagnetic radius of the
nucleus, $Q^2 > 1/R^2$. Therefore, if $Q^2_{\max}\gg 1/R^2$, one can simply
extend the upper limit of the integration in
Eq.~(\ref{eq_photon_number_integral}) to infinity and obtain:
\begin{eqnarray}
\label{eq_photon_number_integral_1}
\dd n_T(\omega_L) =
\frac{Z_1^2\alpha}{\pi}\,\frac{\dd\omega_L}{\omega_L}\,
g\left(\omega_L\,R/\gamma_L\right) \, ,
\end{eqnarray}
where
\begin{equation}
\label{eq_function_g}
g(x) = \int_{x^2}^{\infty} \frac{{\rm d}y}{y} \,\left( 1- \frac{x^2}{y}
\right) \, F^2(y/R^2) \,
\end{equation}
In particular, it was found that for small values of $x =
\omega_L\,R/\gamma_L$, which correspond to the range of parameters
considered in the present article, $g(x)$ can be approximated with
an accuracy better than 1\% as
\begin{equation}
   \label{eq_g_function_approximation}
   g(x) = \ln \frac{1}{x^2} - C_0 \, ,
\end{equation}
with $C_0=0.163$ for Pb and $C_0=0.166$ for Au.

Equations~(\ref{eq_photon_number_integral_1}) and
(\ref{eq_g_function_approximation}) provide an approximation to
the photon number $\dd n_T(\omega_L)$ in the regime where
$Q^2_{\max}\gg 1/R^2$. A simple estimate of the $\dd n_T(\omega_L)$
can be obtained also for  the $Q^2_{\max} \ll 1/R^2$. In the
latter case, the nucleus can be treated as pointlike with
the $F(Q^2) = 1$. By inserting this form factor into
Eq.~(\ref{eq_photon_number_integral}) we find:
\begin{equation}
\label{eq_photon_number_integral_2}
\dd n_T(\omega_L) =
\frac{Z_1^2 \, \alpha}{\pi}\,\frac{\dd\omega_L}{\omega_L}\,
\left[\ln\left( \frac{\gamma^2_L\,Q^2_{\max}}{\omega^2_L} \right) -1\right]\,.
\end{equation}

Having clarified the behavior of the photon numbers $\dd n_{T,S}$,
we now turn to the question of how the cross sections
$\dd \sigma^{T,S}_{\gamma* Z}(\omega_L, Q^2)$, which also enter
Eq.~(\ref{eq_EPA_cross_section}), depend on the virtuality $Q^2$.
In order to perform this analysis, we consider pair
production in the fields of virtual and real photons. This
$\gamma^* \gamma \to e^+ e^-$ process is known~\cite{MeH98} to
exhibit very similar behavior as
(\ref{eq_EPA_pair_production}), but can be described by simple
analytic expressions given in Appendix E of
Ref.~\cite{BuG75}. Namely, while the pair production cross sections
$\sigma_{\gamma^*\gamma}^{T,S}(W^2,Q^2)$ drop drastically at
$Q^2\gg W^2$, the following estimates are valid for smaller values $Q^2 \ll W^2$,
\begin{eqnarray}
\label{eq_cross_sections_Q_behavior}
\sigma_{\gamma^*\gamma}^{S} &\lesssim&
\frac{Q^2}{W^2}\,\sigma_{\gamma\gamma} \,,
\nonumber\\
\sigma_{\gamma^*\gamma}^{T} &=& \sigma_{\gamma\gamma} \,
\left[ 1 + {\cal O}\left(\frac{Q^2}{W^2}\right) \right] \,,
\end{eqnarray}
where $\sigma_{\gamma\gamma} = \sigma_{\gamma\gamma}(W^2)$ describes the $e^+ e^-$ pair
production by two \textit{real} photons.

We are now ready to sum up results of the above analysis and to evaluate the
cross sections for the bound-free pair production.  For the photon virtualities
in the interval $[Q^2_{\min}, Q^2_{\max}]$ (see discussion of
Eq.~(\ref{eq_virtuality_minimum})) we can neglect the contribution of the
scalar photon and approximate the cross section of the pair production by the
transverse photon ${\dd}\sigma^{T}_{\gamma* Z}$ by its value
${\dd}\sigma^{T}_{\gamma Z}$ on the mass shell (i.e. when $Q^2$ = 0). Under
these assumptions, Eq.~(\ref{eq_EPA_cross_section}) simplifies to
\begin{equation}
\label{eq_EPA_cross_section_final}
\dd \sigma^{\rm EPA}_{ZZ} =
\dd n_T(\omega_L)\, \dd \sigma_{\gamma Z}(\omega_L, p_{+\perp}) \, .
\end{equation}
The accuracy of such an approximation should be verified for two kinematic regimes.
If $W^2 \gg 1/R^2$, which corresponds to the experimental conditions
of an LHC detector, the photon's virtuality is restricted by
the nuclear form factor to
\begin{equation}
\label{eq_virtuality_restriction}
Q^2 \lesssim \frac{1}{R^2} \ll W^2
\end{equation}
and we should use the
expressions~(\ref{eq_photon_number_integral})--(\ref{eq_g_function_approximation})
for the number of the equivalent photons $\dd n_T(\omega_L)$.
>From Eq.~(\ref{eq_g_function_approximation}),
we obtain the large Weizs\"acker--Williams logarithm
$\ell = \ln\left[\gamma^2_L/(R\omega_L)^2\right]$ that originates from the
integral
\begin{equation}
\label{omited}
\int_{Q^2_{\min}}^{1/R^2} \frac{\dd Q^2}{Q^2}\,.
\end{equation}
On the other hand, the omitted terms are small,
\begin{equation}
\label{omited2}
\sim \int_{Q^2_{\min}}^{1/R^2} \frac{\dd Q^2}{Q^2}\;
\frac{Q^2}{W^2}\sim \frac{1}{R^2 \, Q^2} \ll 1\,,
\end{equation}
and do not exhibit logarithmic enhancement. Therefore, the accuracy
of the EPA as determined by the relative order of the neglected
terms is determined by the factor
\begin{equation}
\label{eq_accuracy_estimate1}
\eta_1 \sim \frac{1}{(RW)^2 \,\ell} <  1\,\% \, .
\end{equation}

For the total cross section, the main region
corresponds to $W^2 \sim 4m^2 \ll 1/R^2$, and we should use the
expression~(\ref{eq_photon_number_integral_2}) for the number of
equivalent photons $\dd n_T(\omega_L)$. It contains a large
Weizs\"acker--Williams logarithm $\sim \ln\gamma^2_L$. On the
other hand, the omitted items are of the order of unity,
\begin{equation}
\label{omited3}
\sim \int_{Q^2_{\min}}^{4m^2} \frac{\dd Q^2}{Q^2}\;
\frac{Q^2}{W^2}\sim  1\,,
\end{equation}
and again have no logarithmic enhancement. Therefore, the accuracy
of EPA in this case is of the order of
\begin{equation}
\label{eq_accuracy_estimate2}
\eta_2 \sim \frac{1}{\ln(\gamma^2_L)} \sim  3\,\% \, .
\end{equation}

%
% Bound--free pair photo--production in Sauter approximation
%
\subsection{Bound--free pair photo--production in the Sauter approximation}
\label{Sub_sec_SLA_differential}

Finally,
as seen from Eq.~(\ref{eq_EPA_cross_section_final}), the
computation of $\dd\sigma^{\rm EPA}_{ZZ}$ within the EPA can be
traced back to the differential cross section of the
bound--free pair production following (real rather than virtual)
photon impact on a bare nucleus:
\begin{equation}
\label{eq_photo_production}
\gamma +Z \to e^+ + (Z+e^-)_{1s} \, .
\end{equation}
During the past two decades, this process has been studied in detail
with a special emphasis on high--$Z$ ions. The
total as well as differential cross sections have been evaluated
within a relativistic framework in Refs.~\cite{AgS97,ArS12}. The
fully relativistic calculations, based on the partial--wave
representation of the Dirac continuum states, were found to
provide accurate predictions in the near--threshold region but
face well--known problems connected with the slow convergence of
the multipole expansions when the photon energy increases. Some
approximate methods have to be used, therefore, in order to calculate the
cross sections of the process (\ref{eq_photo_production}) for very
high photon energies that correspond to large transfer momenta in
ion--ion collisions of the type~(\ref{eq_process}). In the present work, the
calculations are based on the Sauter approximation (SA) which was
originally derived for the atomic photoeffect by assuming
ultra--relativistic electron energies and disregarding terms of
relative order $\alpha^2 Z^2$ in an $\alpha Z$--expansion of the
transition amplitude \cite{Sau31,FaM59,PrA73}. The general
formulas of such an approximation, when applied to the
bound--free pair photo-production, are obtained below, while their
validity for the high--$Z$ is discussed later in Section
\ref{Sub_sec_differential_calculations}.

The differential cross section for the process~(\ref{eq_photo_production})
can easily be deduced from
Sauter's relativistic formula for the photoeffect
or photorecombination, in the notation
of Eq.~(\ref{eq_photo_production}) as given, for example,
in Eq.~(57.8) of Ref.~\cite{BeL94} upon replacing {the electron
four--momentum $p$  in the photoeffect with the momentum $(-p_+)$
of the outgoing positron in the bound-free photo-production
(crossing symmetry). More
specifically, this implies the following changes:
\begin{eqnarray}
\label{eq_Sauter_transformation}
& &v = \frac{\sqrt{\gamma^2-1}}{\gamma} \to -\frac{\sqrt{\gamma_L^2-1}}{\gamma_L}
= -v_+ \, , \nonumber \\
& & 1-v^2 \to \frac{1}{\gamma_L^2},\;\; 1-\sqrt{1-v^2} \to \frac{\gamma_L+1}{\gamma_L},
\nonumber \\
& & 1-v\cos\theta  \to 1-v_+\cos\vartheta_+\,,
\end{eqnarray}
where $v_+$ and $\vartheta_+$ are the velocity and the scattering
angle of the positron in the nuclear rest frame, and
$\omega_L = (\gamma_L+1)\,m$ is the photon energy. Performing
these substitutions and averaging over the incident photon
polarization, we find
\begin{eqnarray}
\label{eq_Sauter_formula_diff}
\frac{\dd\sigma^{\rm SA}_{\gamma Z}}{d\Omega_+} &=&
\frac{Z^5\alpha^6}{m^2} \frac{v_+ \sin^2{\vartheta_+}}
{\left(\gamma_L+1\right)^4 \left(1-v_+\cos{\vartheta_+} \right)^4} \nonumber \\
& & \hspace*{-1cm} \times \left[ v^2_+ (\gamma_L+2)\left(1-v_+\cos{\vartheta_+}\right)
-2\frac{\gamma_L-1}{\gamma^3_L} \right] \, ,
\end{eqnarray}
where, moreover, an additional factor two was included to account
for the the different statistical weights of leptons in the
initial and final states of the photo-effect and the pair
photo-production.

Equation~(\ref{eq_Sauter_formula_diff}), obtained within the Sauter
approximation, describes the emission pattern of positrons
created in the process (\ref{eq_photo_production}). Integrating
this expression over the positron angles $\vartheta_+$, we find the
\textit{total} cross section of pair photo--production,
\begin{eqnarray}
\label{eq_Sauter_formula_total}
\sigma^{\rm SA}_{\gamma Z} = 4\pi \,\frac{Z^5\alpha^6}{m^2} \,G(\gamma_L) \, ,
\end{eqnarray}
where the function $G(\gamma_L)$ is given by:
\begin{eqnarray}
\label{eq_G_large_function}
G(x) &=& \frac{\sqrt{x^2-1}}{(x +1)^4}\,
\Bigg[x^2 + \frac{2}{3} \, x +\frac{4}{3} \nonumber \\
&-& \frac{x+2}{\sqrt{x^2-1}}\,\ln\left(x+\sqrt{x^2-1} \right) \Bigg] \, .
\end{eqnarray}
One may note that this expression coincides with Eq.~(52) from
Ref.~\cite{AgS97} upon the replacement $E_+ = - \gamma_L$.
In the high--energy limit, $\gamma_L\gg 1$, Eq.~(\ref{eq_Sauter_formula_total}) reads
\begin{equation}
\label{eq_Sauter_formula_total_assymptotic}
\sigma^{\rm SA}_{\gamma Z} = 4\pi \,\frac{Z^5\alpha^6}{m^2\gamma_L} \,,
\end{equation}
where the dominant contribution is from the
first term in square brackets in Eq.~(\ref{eq_G_large_function}).

%
% Figure 2
%
\begin{figure}[t!]
\vspace*{0.5cm}
\begin{center}
\includegraphics[width=7cm]{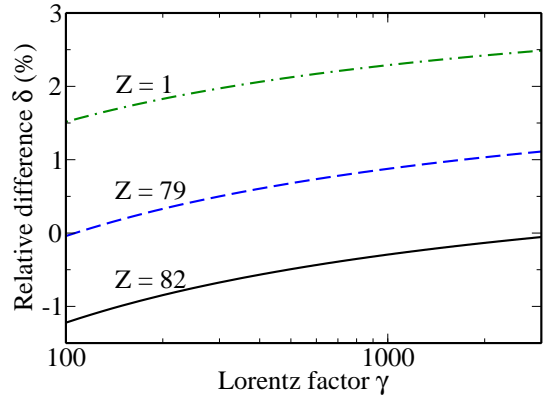}
\end{center}
\caption{(Color online) The relative difference (\ref{eq_relative_difference})
between rigorous relativistic results and predictions of
the scaled EPA--Sauter approximation
(\ref{eq_cross_sections_f_function_1}) for the total cross
sections of bound--free pair production. Calculations are
performed for the energetic collisions between two protons
(dash--and--dotted line) as well as bare gold Au$^{79+}$ (dashed
line) and lead Pb$^{82+}$ (solid line) ions. The EPA--Sauter cross
section is given by Eqs.~(\ref{eq_total_final_2}) and
(\ref{eq_A_B_our}) derived using a value of $Q^2_{\max}=4m^2$.}
\label{Fig2}
\end{figure}

%
% Evaluation of total and differential cross sections
%
\subsection{Evaluation of total and differential cross sections}
\label{Sub_sec_cross_sections}

In the two previous sections, we have shown how the cross section
for the process~(\ref{eq_EPA_pair_production}) can be expressed in
terms of pair photo--production cross sections and we have
evaluated the latter ones within the framework of the Sauter
approximation. Now we are ready to conclude this analysis and to
derive the final expressions that characterize the positron
emission accompanying bound--free pair production in
ultra--relativistic collisions between two ions of equal charge:
\begin{equation}
   Z_1 = Z_2 = Z \, .
\end{equation}
For the high--energy and large--momentum transfer regime, where $\gamma_L m
\gg p_{+ \perp} = m \gamma_L \vartheta_+ \gg m$, the differential
cross section (\ref{eq_Sauter_formula_diff}) simplifies to:
\begin{equation}
\label{eq_Sauter_diff_asymptotic}
\dd\sigma^{\rm SA}_{\gamma Z}(\omega_L, p_{+\perp}) =
16\pi \, \frac{Z^5\alpha^6}{m^2}\;
\frac{m}{\omega_L}\;\frac{m^2\, \dd p_{+\perp}}{p_{+\perp}^3}\,.
\end{equation}
Inserting this expression into
Eq.~(\ref{eq_EPA_cross_section_final}), and using the photon
number~(\ref{eq_photon_number_integral_1}), we obtain the cross
section
\begin{eqnarray}
\label{eq_diff_cross_section_final}
\dd\sigma^{\rm SA}_{ZZ} &=& 16 \,
\frac{\left(Z\alpha\right)^7}{m^2} \,
\frac{m}{\omega_L} \,g\left(\frac{\omega_L\,R}{\gamma_L}\right) \;
\frac{\dd \omega_L}{\omega_L}\;
\frac{m^2 \dd p_{+\perp}}{\left(p_{+\perp}\right)^3}
\\[0.2cm]
&=&8\,\frac{\left(Z\alpha\right)^7}{m^2}\,
\frac{m^3\,{\rm e}^{y_+}}{\gamma\left(p_{+\perp}\right)^4} \,
g\left(\frac{p_{+\perp}R\,{\rm e}^{-y_+}}{2\gamma}\right)
\dd y_+\,\dd p_{+\perp} \,,
\nonumber
\end{eqnarray}
which can be used for an estimate of the positron yield in
LHC experiments. The function
$g$ is defined in Eqs.~(\ref{eq_photon_number_integral_1})
and~(\ref{eq_function_g}).
However, since the central detector in such an
experiment observes the positrons emitted in a wide range of
angles, one has to integrate
Eq.~(\ref{eq_diff_cross_section_final}) over the rapidity $y_+$ in
the interval (\ref{eq_rapidity_restriction}) and over the momentum
$p_{+\perp}\geq p_{\min}$ to get the \textit{partial} cross
section $\Delta\sigma^{\rm SA}_{ZZ}$,
which is relevant to experimental observation:
\begin{equation}
\label{eq_partial_cross_section_final}
\Delta\sigma^{\rm SA}_{ZZ} \approx \frac{16}{3}
\,\frac{\left(Z\alpha\right)^7}{m^2} \,
\frac{\ee^{y_{\min}}}{\gamma}\, \left(\frac{m}{p_{\min}}\right)^3\, L \,.
\end{equation}
The parameter $L$ is given by
\begin{eqnarray}
\label{eq_L_formula}
L &=&
\left[ 2\ln\left( {\frac{\gamma }{R \, p_{\min}}} \right)  +
2 \, y_{\min} - 1.44 \right] \,
\left(1-{\rm e}^{-2y_{\min}}\right)
\nonumber \\[0.2cm]
&& + 4 y_{\min} \, \ee^{-2y_{\min}} \, .
\end{eqnarray}

In addition to the partial positron yield
(\ref{eq_partial_cross_section_final}), we can also employ the
Sauter approximation (\ref{eq_Sauter_formula_total}) in order
to estimate the total bound-free pair production in
ultra-relativistic nuclear collisions. We use
Eq.~(\ref{eq_EPA_cross_section_final}) together with the number of
equivalent photons (\ref{eq_photon_number_integral_2}) where we
adopt $Q^2_{\max}=4m^2$,
and obtain (upon integration over the
photon energy):
\begin{eqnarray}
\label{eq_total_final_1}
\sigma^{\rm EPA}_{ZZ} &=&
\int_{2m}^{\infty} \dd n_T(\omega_L)\,
\sigma^{\rm SA}_{\gamma Z}(\omega_L) \nonumber \\
&=& 4\, \frac{(Z\alpha)^7}{m^2}\,
\left\{a\left[\ln(4\gamma^2_L)-1\right]-b \right\}\,,
\end{eqnarray}
where the parameters $a$ and $b$ can be evaluated analytically,
\begin{eqnarray}
\label{eq_a_b_parameters}
a &=& \int_1^{\infty} G(x)\, \frac{{\rm d}x}{x+1}=\frac{137}{630}=0.2175 \, ,
\nonumber \\[0.2cm]
b &=& \int_1^{\infty} G(x)\,\ln{(x+1)}\, \frac{{\rm d}x}{x+1} \nonumber \\[0.1cm]
&=& \frac{57\,707-19\,180\,\ln2}{88\,200}= 0.5035 \, .
\end{eqnarray}
By taking into account that $\gamma_L$ is defined in terms of the
nuclear Lorentz factors as $\gamma_L = 2 \gamma^2 -1$ we obtain
from Eq.~(\ref{eq_total_final_1}) the well--known scaling law:
\begin{equation}
\label{eq_total_final_2}
\sigma^{\rm EPA}_{ZZ} = A \ln\gamma - B \, ,
\end{equation}
where $A$ and $B$ are predicted to be independent of $\gamma$
\cite{BaR91,BaR93,KrV98}. The values of these two parameters, as
estimated within the equivalent photon approximation, depend on
the particular choice of the maximal photon virtuality
$Q^2_{\max}$. For example, based on our estimate $Q^2_{\max}=4m^2$
which was recommended in Ref.~\cite{MeH98}, we find:
\begin{eqnarray}
\label{eq_A_B_our}
A &=& 16\frac{(Z\alpha)^7}{m^2}\,a =
3.479 \, \frac{(Z\alpha)^7}{m^2}=5.72\,Z^7\;\mbox{pb}\,,
\nonumber \\[0.2cm]
B &=& 8\frac{(Z\alpha)^7}{m^2}\,
\left[b-a(\ln4-0.5)\right] = 2.49 \, \frac{(Z\alpha)^7}{m^2}
\nonumber \\
&=& 4.09\,Z^7\;\mbox{pb}\,,
\end{eqnarray}
which are in the picobarn range (pb).
By contrast, the different value $Q^2_{\max} = 1.26 \, m^2$ adopted
in Ref.~\cite{Ast07} yields
\begin{equation}
\label{eq_A_B_Aste}
A=3.479 \, \frac{(Z\alpha)^7}{m^2} \,,
\qquad B = 3.49 \, \frac{(Z\alpha)^7}{m^2}\,.
\end{equation}
As seen from Eqs.~(\ref{eq_A_B_our}) and (\ref{eq_A_B_Aste}),
while the parameter $A$ is not sensitive to the change of
$Q^2_{\max}$, the parameter $B$ varies by about 40 \% which is
quite remarkable but still within the accuracy of the EPA.

%
% Figure 3
%
\begin{figure}[t!]
\vspace*{0.7cm}
\begin{center}
\includegraphics[width=7cm]{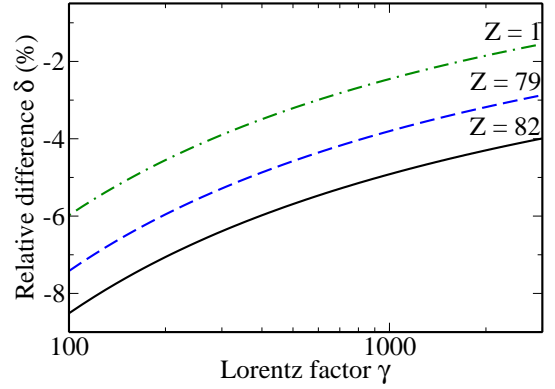}
\end{center}
\caption{(Color online) The same as in Fig.~\ref{Fig2}, but the EPA--Sauter cross
section is given by Eqs.~(\ref{eq_total_final_2}) and
(\ref{eq_A_B_Aste}) employing a value of $Q^2_{\max}=1.26 \, m^2$.}
\label{Fig3}
\end{figure}

%
% Results and discussion
%
\section{Results and discussion}
\label{Sec_Results}

%
% Total cross sections
%
\subsection{Total cross sections}
\label{Sub_sec_total_calculations}

Although the main goal of the present paper is to provide reliable
estimates for the positron yield as might be measured by the
central detector of an LHC experiment, we start calculations
with the total pair production cross sections. Their analysis,
performed for different nuclear charges and various collision
energies, can help to better define the range of validity for our
EPA--Sauter model. In order to start with such an analysis, let us
recall that the high--energy limit of the Sauter photo--production
cross section (\ref{eq_Sauter_formula_total_assymptotic}) is known
to differ from the exact results by some factor:
\begin{equation}
\label{eq_cross_sections_f_function}
\sigma^{\rm exact}_{\gamma Z} =
f(Z) \, \sigma^{\rm SA}_{\gamma Z} \, , \qquad \omega_L \to \infty \, ,
\end{equation}
where $f(Z)$ is a decreasing function of the nuclear charge
\cite{Pra60,MiS93,AsH94}. Values for $f(Z)$ are obtained in
Ref.~\cite{AgS97} based on rigorous relativistic calculations. For
example, for the $e^+e^-$ production in which the electron is
captured into the ground state of an initially bare ion, these
calculations predict $f(Z) = 0.971, 0.222, 0.216$ and $0.196$ for
$Z = 1, 79, 82$ and $92$, respectively.

In the following we shall apply the factor $f(Z)$ to improve
also the accuracy of approximations
(\ref{eq_partial_cross_section_final}) and
(\ref{eq_total_final_2}) that describe the bound--free pair
production in heavy ion collisions. For the total cross section of
this process, for example, we make the replacement:
\begin{equation}
\label{eq_cross_sections_f_function_1}
\sigma^{\rm EPA}_{ZZ} \to f(Z) \cdot
\sigma^{\rm EPA}_{ZZ} = f(Z) \, \left( A \ln\gamma - B \right)\, .
\end{equation}
As seen from Fig.~\ref{Fig2}, where the relative difference
\begin{equation}
\label{eq_relative_difference}
\delta=\frac{f(Z)\cdot\sigma^{\rm EPA}_{ZZ}-
\sigma^{\rm exact}_{ZZ}}{\sigma^{\rm exact}_{ZZ}}
\end{equation}
between the scaled EPA--Sauter cross section and exact results
$\sigma^{\rm exact}_{ZZ}$ from Ref.~\cite{MeH01} is displayed,
such a replacement leads to a very good agreement between the
predictions of the two theories (EPA--Sauter
and rigorous relativistic calculations~\cite{MeH01}). In
particular, for energetic collisions of two gold (dashed line) and
led (solid line) ions the relative difference
$\delta$ does not exceed 1.5\,\% for
Lorentz parameters in the range 100 $\le \gamma \le$ 3000. It is
worth mentioning that a slightly worse performance of the
EPA--Sauter approximation can be observed if in place of the
photon virtuality $Q^2_{\max} = 4m^2$, used to derive values
(\ref{eq_A_B_our}), the $Q^2_{\max} = 1.26 \, m^2$ from
Ref.~\cite{Ast07} is adopted. For the latter case, the relative
difference between the exact and EPA predictions may reach
$6\div12 \, \%$ in the low--energy domain (see Fig.~\ref{Fig3}).

%
% Differential cross sections
%
\subsection{Differential cross sections}
\label{Sub_sec_differential_calculations}

Having briefly discussed computations of the total
bound-free pair production cross sections,
we now estimate the \textit{differential} pair production
probabilities. In order to assess the reliability of these
predictions, let us first verify the accuracy of
Eq.~(\ref{eq_Sauter_diff_asymptotic}) for $\dd\sigma^{\rm SA}_{\gamma Z}$,
and Eq.~(\ref{eq_diff_cross_section_final}) for $\dd\sigma^{\rm SA}_{ZZ}$.
These formulates are based on the Sauter approximation
(\ref{eq_Sauter_formula_diff}) for the $e^+e^-$
photo--production. Sin\-ce this approximation is correct only
in the leading order of $\alpha Z$, its validity in the high--$Z$ domain has to be examined.
To this end, in Fig.~\ref{Fig4}, we display the differential cross
sections $\dd \sigma^{\rm exact}_{\gamma Z}/\dd \Omega_+$ of the process
(\ref{eq_photo_production}) for the photon collision with bare led
ions Pb$^{82+}$ and emitted positron energies $m\gamma_L$ with
$\gamma_L$= 5, 10, 15 and 25. Results of rigorous relativistic
theory, which employs the standard partial-wave decomposition of
continuum Dirac wavefunctions~\cite{ArS12}, are compared
with scaled predictions of Eq.~(\ref{eq_Sauter_formula_diff}),
obtained as follows,
\begin{equation}
\label{eq_Sauter_formula_diff_scaling}
\frac{\dd \sigma^{\rm SA}_{\gamma Z}}{\dd\Omega_+} \to f(Z) \cdot
\frac{\dd\sigma^{\rm SA}_{\gamma Z}}{\dd\Omega_+} \, ,
\end{equation}
where $f(Z)$ is the same factor as employed in
Eq.~(\ref{eq_cross_sections_f_function}). One may note that the
previously observed, good
performance of this scaling for the {\em total} rates (the main
contribution to which comes from forward positron emission,
$\vartheta_+ \lesssim 1/ \gamma_L$), does not automatically
guarantee accurate results for the \textit{differential} cross
sections, especially in the region of larger emission
angles. Fortunately,
the results presented in Fig.~\ref{Fig4} clearly indicate
the validity of this scaling.
Namely, even though the (scaled) Sauter approximation can
significantly underestimate exact results for forward as
well as backward positron emission, the range of angles, over
which the two theories differ significantly, shrinks with
increasing positron energy. For example, while for
$\gamma_L = 5$, the Sauter approximation reproduces the exact
results with the accuracy of about 5 \% only for the angles
$50^\circ < \vartheta_+ < 60^\circ$, this interval increases
to $2^\circ < \vartheta_+ < 120^\circ$ for $\gamma_L= 25$.
These findings justify the application of the Sauter
approximation together with the scaling
(\ref{eq_Sauter_diff_asymptotic}) for the analysis of the pair
production in ultra--relativistic heavy ion collisions at an
LHC experiment in which large positron transverse momenta,
$p_{+\perp}$, will be monitored.

%
% Figure 4
%
\begin{figure*}[t!]
\vspace*{1.2cm}
\begin{center}
\includegraphics[width=14cm]{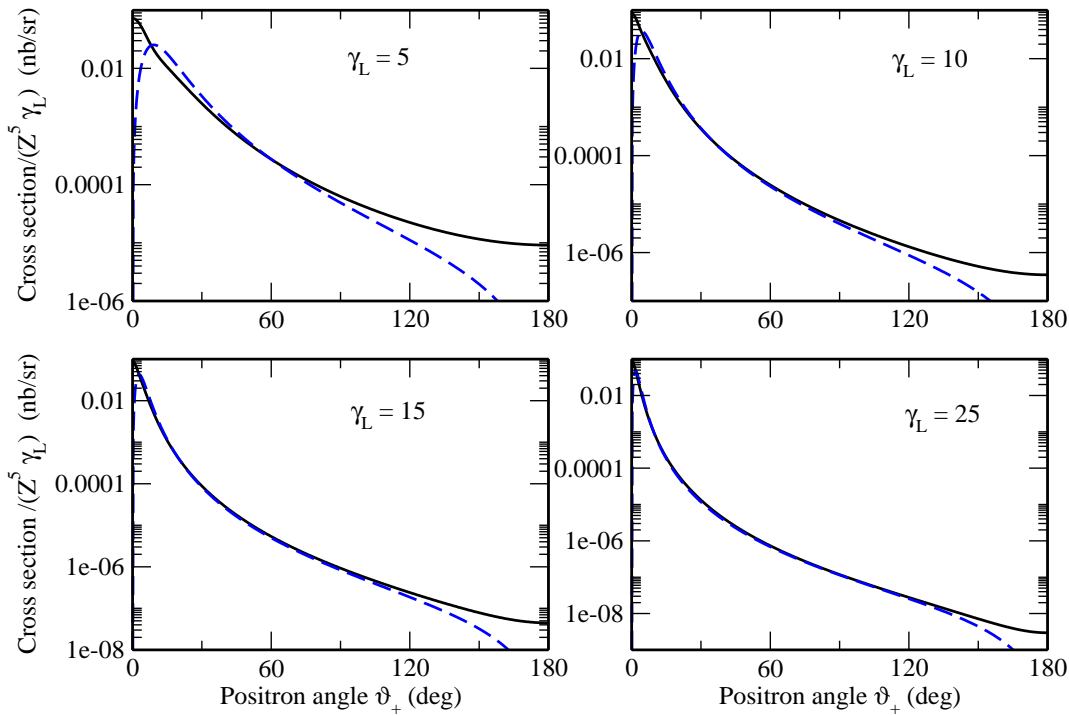}
\end{center}
\caption{(Color online) The differential cross section
$(Z^5\,\gamma_L)^{-1}\,\dd\sigma_{\gamma Z}/\dd\Omega_+$ (in
nanobarn/steradian) of the pair production in the collision of
photon with bare lead Pb$^{82+}$ ion accompanied by the electron
capture into the ground ionic state. Results of the rigorous
relativistic calculations (solid line) are compared to the
predictions of the scaled Sauter approximation (dashed line).
Calculations have been performed in the nucleus rest frame and
four positron energies $m\gamma_L$ with $\gamma_L = 5, 10, 15$ and
$25$.}
\label{Fig4}
\end{figure*}

Besides the direct numerical ``proof'', yet another confirmation of the
scaling conjecture~(\ref{eq_Sauter_formula_diff_scaling}) was recently
received. Namely, the asymptotic behavior of the pair--production
differential cross section in the region $m^2\ll p^2_{+\perp}\ll 2m\omega_L$
was derived for arbitrary $Z\alpha$ in Ref.~\cite{Mil12}. This asymptotic
reads as
\begin{eqnarray}
\label{eq_exact_diff_asymptotic}
&&\dd \sigma^{\rm exact}_{\gamma Z}(\omega_L, p_{+\perp}) =
f^{\rm asymp}(Z)\, 16\pi
\frac{Z^5\alpha^6}{m^2}\; \frac{m}{\omega_L}\;\frac{m^2\,
\dd p_{+\perp}}{p_{+\perp}^3} \,,
\nonumber\\[0.2cm]
&& f^{\rm asymp}(Z)=
\frac{2\,(2\eta)^{2\tilde\gamma-2}}{\Gamma(2\tilde\gamma+1)}\,\,
|\Gamma(\tilde\gamma-{\rm i} \eta)|^3\,{\rm
e}^{-2\eta\arccos{\eta}}\,,
\end{eqnarray}
where $\eta=Z\alpha$ and $\tilde\gamma=\sqrt{1-\eta^2}$. One obtains from this expression $f^{\rm asymp}(Z)$ = 0.297, 0.287 and 0.257
for $Z$ = 79, 82 and 92. Therefore, our assumptions about scaled factor $f(Z)$ are by about 25 \% smaller then the corresponding factor $f^{\rm asymp}(Z)$ for the same values of $Z$. We note that such a discrepancy is comparable to the (expected) experimental error and the overall uncertainty of our model that arises not only from the use of the Sauter and EPA approximations but  also from neglecting the electron capture into excited ionic states, an effect which will be briefly addressed later.

Based on the accuracy analysis of the scaled Sauter approximation
(\ref{eq_Sauter_diff_asymptotic}) for $\dd\sigma^{\rm SA}_{\gamma Z}$
and (\ref{eq_Sauter_formula_diff_scaling}) for $\dd\sigma^{\rm SA}_{ZZ}$,
we are now in the position to compute the positron yield. This yield is defined by the
partial cross section given in
Eq.~(\ref{eq_partial_cross_section_final}), which has to be scaled
properly,
\begin{equation}
\label{eq_partial_cross_section_final_scaled}
\Delta\sigma^{\rm SA}_{ZZ} \to 2 \, f(Z) \, \Delta\sigma^{\rm SA}_{ZZ} \,.
\end{equation}
The factor two accounts for the fact that the electron can
be captured by either of two nuclei. Assuming a maximum
luminosity $\mathcal{L} = 10^{27} \, {\rm s}^{-1} \, {\rm cm}^{-2}$
which can be reached in Pb--Pb collision experiments at CERN, and
using the corresponding sets of kinematic parameters
$\{y_{\rm min}, p_{\rm min} \}$ (see Sect.~\ref{Sec_parameters}), we finally
obtain from Eqs.~(\ref{eq_partial_cross_section_final}) and
(\ref{eq_partial_cross_section_final_scaled}) that
{\bf (i)} only one event per 67 days is likely to be observed for the
high--transverse--momentum scenario (\ref{eq_scenario_A0}),
whereas {\bf (ii)} about 16 counts per hour are within the
acceptance range of the detector within the
scenario (\ref{eq_scenario_B0}).

One may note that these estimates were obtained based on the
assumption that the electron is captured into the ground ionic
states. The $e^+ e^-$ pair production accompanied by the
formation of excited hydrogen--like ions may slightly increase
these predictions. However, Refs.~\cite{AgS97,LeM04}
indicate that the effect of the excited state
electron recombination shall not exceed $20 \div 25\,\%$.

%
% Summary and outlook
%
\section{Summary and outlook}
\label{Sec_summary}

In conclusion, we investigate bound--free pair
production in ultra--relativistic heavy ion collisions. Special
emphasis in our study is placed on the emission of positrons with
large transverse momenta which can be observed at the LHC collider. In order to estimate the
yield of such positrons, an approximate method, based on the
equivalent photon approximation and the Sauter theory, has been
laid out. Within the EPA--Sauter approach, we derive a simple
analytical expression for the differential cross section of the
pair production process (\ref{eq_process}). Based on this
expression, calculations have been performed for the collisions
between two bare lead ions Pb$^{82+}$ moving towards each other
with the Lorentz factor $\gamma$ = 1500. At this energy, typical
for the LHC, we consider two scenarios that
correspond to different detectors set-ups. While for the first
scenario (\ref{eq_scenario_A0}) the number of events is too small
to be measured,  the second scenario (\ref{eq_scenario_B0}) looks
rather promising. Our conceptually simple approach allows
us to incorporate the detector set--up into the
final formulas for the partial cross sections relevant
to experimental observation at the LHC. Essentially, we conclude
that in the first scenario~(\ref{eq_scenario_A0}),
bound-free pair production takes place but most of the positrons
escape the detector; they are not sufficiently deflected out of
the beam line to be within the acceptance range of the LHC detectors.

In the present work we have restricted our theoretical analysis to
a \textit{single} $e^+ e^-$ production. In the
ultra--relativistic regime,  however, the ion--ion collisions may
result in the creation of a few electron--positron or even
muon--antimuon pairs. Such a multiple--pair (and heavier lepton)
production now attracts considerable attention as a tool for
exploring the quantum electrodynamics in extremely strong
electromagnetic fields~\cite{AlH97,Bal09}. In heavy ion colliders,
these processes again can be explored by measuring the emitted
positrons and residual few--electron ions in coincidence.
Theoretical predictions, required for preparing and analyzing such
coincidence experiments, may be naturally obtained within the
framework of the EPA--Sauter theory. Investigations along this
line are currently underway and will be presented in a future
paper.

%
% Acknowledgments
%
\section*{Acknowledgments}

We are grateful to R.~Schicker who attracted our attention to this problem and
explained us the details of several LHC experiments where the lepton pair--production
measurements might be feasible. The stimulating discussions
with A.~Milstein and A.~Voitkiv are gratefully acknowledged.
U.D.J.~acknowledges support from the National Science Foundation (PHY-1068547)
and from the National Institute of Standards and Technology (precision
measurement grant).  V.G.S.~is supported by the Russian Foundation for Basic
Research via grants 09-02-00263 and 11-02-00242. A.A.~and A.S.~acknowledge
support from the Helmholtz Gemeinschaft (Nachwuchsgruppe VH--NG--421).

%
%
% ---------------------------------------- References ------------------------------- %
%
%

\end{document}